\documentclass[showpacs,amsmath,amssymb,prl,aps,twocolumn]{revtex4}
\usepackage{graphicx,bm}
\begin{document}
\title {Equilibrium spin currents: Non-Abelian gauge invariance and color diamagnetism in condensed matter}
\author {I. V. Tokatly}
\email{Ilya.Tokatly@physik.uni-erlangen.de}
\affiliation{European Theoretical Spectroscopy Facility (ETSF),\\ Dpto. Fisica Materiales, Universidad del Pais Vasco EHU-UPV, 20018 Donostia, Spain\\
and Moscow Institute of Electronic Technology, Zelenograd, 124498 Russia}

\date{\today}
\begin{abstract}
The spin-orbit (SO) interaction in condensed matter can be described in terms of a non-Abelian potential known in high-energy physics as a color field. I show that a magnetic component of this color field inevitably generates diamagnetic color currents which are just the equilibrium spin currents discussed in a condensed matter context. These dissipationless spin currents thus represent a universal property of systems with SO interaction. In semiconductors with linear SO coupling the spin currents are related to the effective non-Abelian field via Yang-Mills magnetostatics equation. 
\end{abstract}
\maketitle

SO interaction is considered as an important ingredient of spintronics \cite{ZutFabSar2004,Rashba2007_Book} as it allows to control spin degrees of freedom by electric means. Despite an increasing interest and growing number of publications on spin dynamics and spin currents in systems with SO interaction, a few basic questions remain unresolved up to now. In fact, even the very definition of spin currents is still debated \cite{Rashba2007_Book,EngRasHal2007_Book,Shi2006}. The reason for the controversies is that the spin is not conserved in a usual sense if SO interaction is present. The time derivative of the spin density ${\bf s}({\bf r},t)$ can not be represented in form of divergence of a current, but always contains an extra term -- the spin torque. Hence it appears that any redefinition of the current can be compensated by correcting the torque in a way that preserves $\partial_t{\bf s}$. The problem of ambiguity of the spin current was sharpened by Rashba who noticed the presence of spin currents in a thermodynamically equilibrium 2D electron gas with Rashba SO interactions \cite{Rashba2003}. The physical reality of these dissipationless currents has been questioned as they do not accompanied by any spin accumulation, and their very appearance has been attributed to the ambiguity of the spin current concept.

Recently \cite{Sonin2007a,Sonin2007b} Sonin proposed a way to detect the equilibrium spin flows in a "Rashba medium", and argued that there is nothing miraculous or unique in the presence of such equilibrium currents. They are known to exist in many different physical situations, e.~g., spin supercurrents in a superfluid He$^3$ or Meissner currents in superconductors (see. Ref.~\onlinecite{Sonin2007a} and references therein). I would like to add another simple example that, as we will see, is much more relevant -- the diamagnetic currents responsible for the Landau diamagnetism in metals. 

This paper is aimed at completely removing a flavor of ambiguity and mystery in the questions of equilibrium spin flows and the definition of spin currents. The problem is resolved by making a link to non-Abelian gauge theories, and exploiting SU(2) gauge invariance of a many-body theory with SO interactions. The interpretation of SO interaction together with an external magnetic field as components $\bm{\mathcal A}_{\mu}$ of a non-Abelian four-potential is known for many years \cite{FroStu1993}, and it is becoming more and more popular nowadays 
\cite{SO-SU2}. However, the full power of non-Abelian gauge invariance in SO context is by far not explored \cite{note_SpinHall}. In the present work I use this analogy to discover a simple physics behind equilibrium spin currents. We will see that these are nothing but diamagnetic color currents that appear as a response to an effective Yang-Mills magnetic field produced by SO interaction. Due to the gauge invariance they are nonzero only if the field strength $\bm{\mathcal F}_{ij}$ is nonvanishing, which is the case in most real situations. This simple physical picture clearly demonstrates the universality of equilibrium spin currents in matter. They should be generically present in almost any system, like molecules or solids, if the SO coupling is non-negligible. Interestingly, in semiconductors with linear SO coupling the spin current is related to the non-Abelian field by a Yang-Mills magnetostatics equation, which makes one more unexpected connection between the condensed matter and high energy physics. The present results also apply to ultracold atomic gases where a background non-Abelian field can be generated optically \cite{Osterloh2005,Ruseckas2005}. 

In general the many-body Hamiltonian with first spin-dependent relativistic corrections can be represented in a form \cite{FroStu1993}
\begin{eqnarray} \nonumber
	H &=& \int d{\bf r}\Big\{\frac{1}{2m}[(i\partial_i + \bm{\mathcal A}_i)\Psi]^{\dagger}[(i\partial_i + \bm{\mathcal A}_i)\Psi] 
	 - \Psi^{\dagger}\bm{\mathcal A}_0\Psi \\
	 &+& U\Psi^{\dagger}\Psi + \frac{1}{2}\int d{\bf r'}V_{{\bf r}{\bf r'}}
	 \Psi^{\dagger}({\bf r})\Psi^{\dagger}({\bf r'})\Psi({\bf r'})\Psi({\bf r})\Big\}
	 \label{1}
\end{eqnarray}
where $\Psi^{\dagger}=(\psi^{\dagger}_{\uparrow},\psi^{\dagger}_{\downarrow})$ is a two-component fermionic field, $V_{{\bf r}{\bf r'}}$ is the interparticle interaction, and $U$ is the external scalar potential \cite{note_A}. The components $\bm{\mathcal A}_{\mu}$ ($\mu = 0,x,y,z$) of the gauge field are 2$\times$2 matrices of the form $\bm{\mathcal A}_{\mu} = {\mathcal A}_{\mu}^a\tau^a$, where $\tau^a = \sigma^a/2$ are the generators of SU(2) (spin-1/2 operators) with the following algebra
$[\tau^a,\tau^b] = i\varepsilon^{abc}\tau^c$, ${\rm tr}\{\tau^a,\tau^b\}= \delta^{ab}$ \cite{note_notation}.

The non-Abelian potential $\bm{\mathcal A}_{\mu}$ captures all spin effects if one makes the following identifications
\begin{equation}
	\bm{\mathcal A}_0 = -\frac{e\hbar}{mc}B^a\tau^a, \quad
	\bm{\mathcal A}_i = \frac{e\hbar}{mc^2}\varepsilon_{ija}E_j\tau^a
	\label{3}
\end{equation}
where $B^a$ are the components of the external magnetic field, and $E_j$ is the electric field produced, for example, by nuclei in molecules or solids. The Hamiltonian of Eq.~(\ref{1}) also covers popular 2D models of semiconductors with linear SO interaction of Rashba and/or Dresselhaus form \cite{Winkler}. In this case the time component of SU(2) potential is still given by ${\mathcal A}_0^a = g\mu_{\rm B}B^a$, but with an appropriate $g$-factor, while the spatial components are defined as follows, $\bm{\mathcal A}_z = 0$,
\begin{equation}
	\bm{\mathcal A}_x = 2m(\beta\tau^x - \alpha\tau^y), \quad
	\bm{\mathcal A}_y = 2m(\alpha\tau^x - \beta\tau^y),
\label{4}
\end{equation}
where $\alpha$ and $\beta$ are the Rashba and Dresselhaus SO coupling constants respectively. 

The beauty of the representation (\ref{1}) is that the corresponding action
\begin{equation}
	S[\Psi,\bm{\mathcal A}_{\mu}] = \int dt(d{\bf r}\Psi^{\dagger}i\partial_t\Psi - H)
\label{5}
\end{equation}
is invariant with respect to local non-Abelian gauge transformations
\begin{equation}
	\Psi \mapsto {\mathcal U} \Psi, \qquad
	\bm{\mathcal A}_{\mu} \mapsto {\mathcal U}\bm{\mathcal A}_{\mu}{\mathcal U}^{-1} -
	i(\partial_{\mu}{\mathcal U}){\mathcal U}^{-1}
	\label{6}
\end{equation}
where ${\mathcal U}=e^{i\theta^a({\bf r},t)\tau^a}$ is an arbitrary SU(2) matrix. The above gauge invariance immediately implies {\em covariant} conservation of a color current, ${\bm J}_{\mu} = J_{\mu}^a\tau^a$, with components $J_{\mu}^a=\delta S/\delta{\mathcal A}_{\mu}^a$:
\begin{equation}
	D_t {\bm J}_0 + D_i{\bm J}_i = 0
	\label{7}
\end{equation}
where $D_{\mu}\cdot= \partial_{\mu}\cdot -i[\bm{\mathcal A}_{\mu},\cdot]$ is a covariant derivative, and
\begin{eqnarray}
  J_{0}^a &=& \frac{\delta S}{\delta{\mathcal A}_0^a} = \Psi^{\dagger}\tau^a\Psi \equiv s^a({\bf r},t)
  \label{8}\\
	J_{i}^a = \frac{\delta S}{\delta{\mathcal A}_i^a} &=&
	 \frac{-i}{2m}[\Psi^{\dagger}\tau^a\partial_i\Psi - (\partial_i\Psi^{\dagger})\tau^a\Psi] -
	\frac{{\mathcal A}_i^a}{4m}\hat{n}
	\label{9}
\end{eqnarray}
Explicitly the covariant conservation law of Eq.~(\ref{7}) reads
$$
\partial_t J_0^a + \varepsilon^{abc}{\mathcal A}_0^bJ_0^c + 
\partial_i J_i^a + \varepsilon^{abc}{\mathcal A}_i^bJ_i^c = 0.
$$
Apparently the second and the fourth terms in this equation violate conservation of the spin $J_0^a=s^a$. The second term causes the spin precession in the U(1) magnetic field ${\mathcal A}_0$. The fourth term is the "internal torque" due to SO interaction. However, the variational definition of the spin four-current, Eqs.~(\ref{8}) and (\ref{9}), based on the gauge invariance, leaves no room for an ambiguity. As soon as we identify the zeroth component, $\delta S/\delta{\mathcal A}_{0}^a$, with the spin density, we are forced to accept that the spatial part, $\delta S/\delta{\mathcal A}_{j}^a$, is the spin current. $J_i^a$ is coupled to ${\mathcal A}_i^a$ in exactly the same fashion as $J_0^a$ is coupled to ${\mathcal A}_0^a$, which is absolutely analogous to the familiar case of the charge four-current coupled to U(1) gauge field. One can also show that $J_i^a$, Eq.~(\ref{9}), is a proper dissipative current conjugated to an effective SU(2) electric field ${\mathcal F}_{0i}^a$ \cite{note_dissipation}. It is worth noting that Eq.~(\ref{9}) coincides with the "natural" definition of the spin current \cite{Rashba2003,Rashba2007_Book,EngRasHal2007_Book}.

Armed with the gauge invariant Hamiltonian and the variational definition of $J_i^a$ we are ready to approach the problem of equilibrium spin currents. SO interaction enters the Hamiltonian as an effective background non-Abelian field. If a magnetic part of this color field is nonzero one naturally expects an orbital response in a form of color diamagnetic currents. These currents, if exist, are given by the derivative of the energy (thermodynamic potential) $E[{\mathcal A}_i^a]=\langle H\rangle$ with respect to ${\mathcal A}_i^a$. Since the energy is gauge invariant it can depend on ${\mathcal A}_i^a$ only via invariants composed of the field strength
\begin{equation}
\bm{\mathcal F}_{ij} = \partial_i\bm{\mathcal A}_{j} - \partial_j\bm{\mathcal A}_{i} 
 - i[\bm{\mathcal A}_{i},\bm{\mathcal A}_{j}]
	\label{10}
\end{equation} 
A particular form of invariants is determined by the symmetry of a particular system.

For the sake of clarity I consider explicitly the case of semiconductors with linear SO coupling of the Rashba-Dresselhaus type. I also assume that the external scalar potential and the usual magnetic field are zero, i.~e. in Eq.~(\ref{1}) $U=0$ and $\bm{\mathcal A}_0=0$. Since in the absence of ${\mathcal A}_i^a$ the system is rotationally invariant, the first SO correction to the energy must be proportional to ${\rm tr}(\bm{\mathcal F}_{ij}\bm{\mathcal F}_{ij})$, i.~e.
\begin{equation}
	E_{\rm SO} = \frac{\lambda}{4}\int d{\bf r}{\mathcal F}_{ij}^a{\mathcal F}_{ij}^a,
	\label{11}
\end{equation}
where $\lambda$ is a constant (on dimensional grounds $\lambda\sim p_F^{D-2}/m$, where $p_F$ is the Fermi momentum and $D>1$ is the dimension of space). Calculation of the current, $J_i^a = -\delta E_{\rm SO}/\delta {\mathcal A}_i^a$, yields
\begin{equation}
	\bm J_j = \lambda D_i\bm{\mathcal F}_{ij} = \lambda \left(\partial_i\bm{\mathcal F}_{ij} 
	- i[\bm{\mathcal A}_{i},\bm{\mathcal F}_{ij}]\right),
\label{12}	
\end{equation}
which is exactly of the form of Yang-Mills magnetostatics equation. Physically the result is very similar to the case of U(1) magnetic field: an external field produces diamagnetic currents aimed at compensating that field. There is, however, an essential difference in the spatial distribution of diamagnetic currents. In U(1) case the currents in the bulk vanish when the magnetic field approaches a constant. Only the integral defining the induced magnetic moment remains finite. In contrast, in the non-Abelian case bulk diamagnetic currents exist even for a constant in space field. The reason is the commutator in the right hand side of Eq.~(\ref{12}). A similar commutator in Eq.~(\ref{10}) gives a nonzero magnetic field even for a space-independent vector potential. Thus, in the case of a homogeneous field (space-independent SO coupling constants), the bulk spin current is given by
\begin{equation}
	\bm J_j = -\lambda[\bm{\mathcal A}_{i},[\bm{\mathcal A}_{i},\bm{\mathcal A}_{j}]]
	\label{13}
\end{equation}
Equation (\ref{13}) shows that in the homogeneous system the spin current is proportional to the third power of non-Abelian potential. This naturally explains why Rashba's equilibrium spin current is proportional to $\alpha^3$ \cite{Rashba2003}. 

The above phenomenology can be confirmed by direct microscopic calculations of the spin current for an exactly solvable model of noninteracting particles. First I consider homogeneous $\bm{\mathcal A}_{i}$. In this case the spin current, Eq.~(\ref{9}), is given by the expression
\begin{equation}
	J_i^a = 
	T\sum_w\sum_{\bf p}{\rm tr}\left[\frac{p_i}{m}\tau^a \hat{G}(\omega,{\bf p})\right]- \frac{n}{4m}{\mathcal A}_i^a
	\label{14}
\end{equation}
where $\omega$ is the fermionic Matsubara frequency, and $n$ is the density of particles. The one particle Green's function $\hat{G}(\omega,{\bf p})$ is defined as follows
 \begin{eqnarray}\nonumber
 &&\hat{G}(\omega,{\bf p})=\left[i\omega + \mu - \frac{1}{2m}(p_j - {\mathcal A}_j^b\tau^b)^2\right]^{-1}\\
&=& \frac{\hat{Z}_{+}({\bf p})}{i\omega +\mu - E_{+}({\bf p})} + 
    \frac{\hat{Z}_{-}({\bf p})}{i\omega +\mu - E_{-}({\bf p})},
\label{15}
\end{eqnarray}
where 
 \begin{eqnarray}
\hat{Z}_{\pm}({\bf p}) &=& \frac{1}{2}
       \left(1 \mp \frac{2p_i{\mathcal A}_i^a\tau^a}{\sqrt{(p_j{\mathcal A}_j^b)(p_k{\mathcal A}_k^b)}}\right),
 \label{16} \\
E_{\pm}({\bf p}) &=& \frac{{\bf p}^2}{2m} + \frac{{\mathcal A}_i^a{\mathcal A}_i^a}{8m} \pm
                  \frac{1}{2m}\sqrt{(p_j{\mathcal A}_j^b)(p_k{\mathcal A}_k^b)}
 \label{17}                         
\end{eqnarray}
After summation over $\omega$ Eq.~(\ref{14}) takes the form
\begin{eqnarray}
J_i^a &=& \frac{1}{2m}\sum_{\bf p}
	\frac{p_i(p_j{\mathcal A}_j^a)}{\sqrt{(p_l{\mathcal A}_l^b)(p_k{\mathcal A}_k^b)}}[n_F(E_{-}) - n_F(E_{+})]
\nonumber \\
      &-& \frac{{\mathcal A}_i^a}{2m}\sum_{\bf p}[n_F(E_{-}) + n_F(E_{+})],
\label{18}      
\end{eqnarray}
where $n_F(E)$ is the Fermi distribution function. In the second term in Eq.~(\ref{18}) the density of particles is represented as $n = \sum_{\bf p}[n_F(E_{-}) + n_F(E_{+})]$. The rest of calculations is straightforward. Assuming as usual that SO coupling is weak, we expand the distribution functions $n_F(E{\pm})$ in terms of $\sqrt{(p_j{\mathcal A}_j^b)(p_k{\mathcal A}_k^b)}$, and keep the first nonvanishing term in Eq.~(\ref{18}). This term is, as expected, of the third order in ${\mathcal A}_j^a$. The final result at zero temperature is the following
\begin{equation}
	J_i^a = \frac{N_F}{24m^2}({\mathcal A}_j^a{\mathcal A}_i^b{\mathcal A}_j^b -
	{\mathcal A}_i^a{\mathcal A}_j^b{\mathcal A}_j^b),
\label{19}	
\end{equation}
where $N_F$ is the density of states at the Fermi level. It is easy to see that the expression in the brackets in Eq.~(\ref{19}) is exactly the double commutator entering the right hand side of Eq.~(\ref{13}). Hence the phenomenological coefficient in the SO energy is $\lambda = N_F/24m^2$. Thus the direct calculations indeed confirm a diamagnetic nature of the equilibrium spin currents. The spin current in the 2D Rashba-Dresselhaus model is obtained by inserting ${\mathcal A}_i^a$ of Eq.~(\ref{4}) and $N_F=m/2\pi$ into Eq.~(\ref{19})
\begin{eqnarray}
	\label{20}
	J_x^y = - J_y^x &=& \frac{m^2}{6\pi}\alpha(\alpha^2 - \beta^2),\\
\label{21}
	J_x^x = - J_y^y &=& \frac{m^2}{6\pi}\beta(\alpha^2 - \beta^2).	
\end{eqnarray}
Setting $\beta=0$ we exactly recover the result by Rashba \cite{Rashba2003}. It is very interesting to realize that the formula for the spin current obtained in Ref.~\onlinecite{Rashba2003} is a hidden form of a covariant curl of the non-Abelian magnetic field! The spin current of Eqs.~(\ref{20}), (\ref{21}) is zero at $\alpha = \pm\beta$. The reason is that the color magnetic field ${\mathcal F}_{xy}^z = 4m^2(\alpha^2-\beta^2)$ vanishes at these special values of SO constants. In the absence of magnetic field there are no diamagnetic currents.

Now I will show that the gauge invariance allows to significantly simplify practical calculations of the spin current using powerful techniques of the linear response theory, in spite of an obvious nonlinearity in ${\mathcal A}_i^a$. Let us first assume that the non-Abelian potential ${\mathcal A}_i^a({\bf r})$ is weak (here we allow for a general inhomogeneity of the potential). The standard linear response theory \cite{GiulianiVignale} yields
\begin{equation}
	J_i^a({\bf r}) = \int d{\bf r}\chi_{ij}^{ab}({\bf r},{\bf r'})\bm{\mathcal A}_j^b({\bf r'}),
	\label{22}
\end{equation}
where the response function is defined by the Kubo formula
\begin{equation}
	\chi_{ij}^{ab}({\bf r},{\bf r'}) = \langle\langle\hat{J}_i^a({\bf r});\hat{J}_j^b({\bf r'})\rangle\rangle - 
	\delta({\bf r}-{\bf r'})\delta_{ij}\delta^{ab}\frac{n}{4m}.
	\label{23}
\end{equation}
The operator $\hat{J}_i^a$ in Eq.~(\ref{23}) is given by the first term in the right hand side of Eq.~(\ref{9}).

Due to the gauge invariance the vector potential ${\mathcal A}_i^a$ can enter Eq.~(\ref{22}) only via the linearized field strength, $\partial_i{\mathcal A}_j^a-\partial_j{\mathcal A}_i^a$, and, possibly, its spatial derivatives, $\partial_i$. By the same token, to recover the formula valid to the first order in the full nonlinear field strength, ${\mathcal F}_{ij}^a$ of Eq.~(\ref{12}), we simply have to make the following replacements 
\begin{equation}
	(\partial_i{\mathcal A}_j^a-\partial_j{\mathcal A}_i^a) \mapsto {\mathcal F}_{ij}^a, \qquad
	\partial_i \mapsto D_i.
	\label{24}
\end{equation}
All other changes are forbidden by the gauge invariance. 

To demonstrate how this prescription works in practice I consider again the example of a homogeneous (in the absence of ${\mathcal A}_i^a({\bf r})$) and noninteracting electron gas with linear SO interaction. In this case Eq.~(\ref{23}) for the spin current response function in the momentum representation simplifies as follows $\chi_{ij}^{ab}({\bf q})=\delta^{ab}\chi_{ij}^{S}({\bf q})$,
\begin{equation}
	\chi_{ij}^{S}({\bf q}) =\sum_{\bf p}p_ip_j
	\frac{n_F(\xi_{\bf{p}-\frac{\bf q}{2}}) - n_F(\xi_{\bf{p}+\frac{\bf q}{2}})}{2m^2(\xi_{\bf{p}+\frac{\bf q}{2}} - \xi_{\bf{p}-\frac{\bf q}{2}})}
 - \delta_{ij}\frac{n}{4m}
 \label{25}
\end{equation}
where $\xi_{\bf p}=\frac{{\bf p}^2}{2m}$. Equation (\ref{25}) coincides, up to a numerical factor, with a static charge current response function $\chi_{ij}({\bf q})$ \cite{GiulianiVignale}. It the standard theory the function $\chi_{ij}({\bf q})$ determines the Landau diamagnetic response to an external magnetic field, which provides us with another confirmation of the diamagnetic nature of the equilibrium spin currents. To calculate the spin current response function we follow the usual route \cite{GiulianiVignale}. In the limit of $q\ll p_F$, which means that SO constants are slowly changing on the scale of $p_F^{-1}$, $\chi_{ij}^{ab}({\bf q})$ takes the form
\begin{equation}
	\nonumber
	\chi_{ij}^{ab}({\bf q}) = \frac{N_F}{24m^2}\delta^{ab}(q_iq_j - q^2\delta_{ij}).
\end{equation}
Transforming this equation from ${\bf q}$- to ${\bf r}$-space, and inserting the result into Eq.~(\ref{22}) we get for the spin current
\begin{equation}
	\label{26}
	J_j^a = \frac{N_F}{24m^2}\partial_i (\partial_i{\mathcal A}_j^a - \partial_j{\mathcal A}_i^a)
\end{equation}
The final step is a substitution of Eq.~(\ref{24}), which yields the final gauge covariant expression valid to the first order in the total non-Abelian magnetic field $\bm{\mathcal F}_{ij}$, Eq.~(\ref{10}),
\begin{equation}\nonumber
	J_j^a = \frac{N_F}{24m^2}D_i{\mathcal F}_{ij}^a.
\end{equation}
Thus the standard linear response theory supplemented with the substitution of Eq.~(\ref{24}) exactly recovers the Yang-Mills form of the spin current, Eq.~(\ref{12}), with the correct coefficient $\lambda=N_F/24m^2$. An obvious advantage of this way is that it allows to straightforwardly include the effects of interaction and external inhomogeneities (for example, impurity scattering). The formalism is also easily transferable to nonequilibrium situations.

Apparently the general conclusion about the nature of equilibrium spin currents is not restricted to the simple Rashba-Dresselhaus model. Such diamagnetic currents necessarily present in any system with SO interaction, provided the effective color magnetic field of Eq.~(\ref{10}) is nonzero. As vanishing  $\bm{\mathcal F}_{ij}$ is actually an exception (like in the case $\alpha=\pm\beta$) it should be possible to find the spin currents in many molecules and solids using available codes for {\em ab initio} electronic structure calculations.

In conclusion I identified the equilibrium spin currents with diamagnetic color currents in the presence of a non-Abelian  field generated by SO coupling. If the particles have a color charge coupled to physical Yang-Mills fields, like in quark-gluon plasma, the color currents would produce a back reaction field to compensate the external one. The absence of such back reaction does not make the equilibrium spin currents less "real". They are as real as pseudo-diamagnetic currents in a rotating reference frame. One of the main outcomes of this work is a conclusion about universality of equilibrium spin currents that should exist in most real systems. The universality makes it especially intriguing to observe and, possibly, to control such currents experimentally. In this respect a proposal by Sonin \cite{Sonin2007b}, connecting spin currents to mechanical deformations, looks especially interesting. The present results show that it is not necessarily to do experiments with a "Rashba medium". Any technologically convenient material with strong SO interaction should demonstrate the same effect. Possibly the color diamagnetic currents can be also observed in trapped atomic gases subjected to a proper configuration of artificial non-Abelian fields \cite{Osterloh2005,Ruseckas2005}.

I am grateful to G. Vignale for valuable comments. This work was supported by the Ikerbasque Foundation.
   

\begin{thebibliography}{22}
\bibitem{note_SpinHall}
\bibinfo{note}{For example, it remains unnoticed that the vanishing static spin
  Hall conductivity in Rashba systems is a direct consequence of the SU(2)
  gauge invariance.}
\bibitem{note_A}
\bibinfo{note}{The usual U(1) external vector potential $A_i$ is not shown in
  Eq.~(\ref{1}). If needed, it is added additively to $\bm{\mathcal A}_i$.}
\bibitem{note_notation}
\bibinfo{note}{The Greek subscripts $\mu,\nu\dots$ label space-time
  coordinates, while the Latin subscripts $i,j,k,\dots$, and superscripts
  $a,b,c,\dots$ are used to label spatial coordinates and the directions in the
  internal spin space, respectively. The summation over repeated indexes is
  assumed throughout.}
\end{thebibliography}

\end{document}